\documentclass[conference]{IEEEtran}
\IEEEoverridecommandlockouts
\usepackage{cite}
\usepackage{amsmath,amssymb,amsfonts}
\usepackage{algorithmic}
\usepackage{graphicx}
\usepackage{textcomp}
\usepackage{xcolor}
\usepackage{epstopdf}
\usepackage[english]{babel}
\usepackage{booktabs}

\begin{document}
\title{ChemMiner: A Large Language Model Agent System for Chemical Literature Data Mining}
\author{
    \IEEEauthorblockN{Kexin Chen$^{1}$, Yuyang Du$^{1*}$, Junyou Li$^{2}$, Hanqun Cao$^{1}$, Menghao Guo$^{2}$, Xilin Dang$^{1}$,  Lanqing Li$^{1,2}$, \\ Jiezhong Qiu$^{3}$, Guangyong Chen$^{3*}$, Pheng Ann Heng$^{1}$}
    \IEEEauthorblockA{$^1$ The Chinese University of Hong Kong, New Territories, Hong Kong SAR}
    \IEEEauthorblockA{$^2$ Zhejiang Lab, Zhejiang University, Hangzhou, China}
    \IEEEauthorblockA{$^3$ Hangzhou Institute of Medicine, Chinese Academy of Sciences, Hangzhou, China}
    \IEEEauthorblockA{\thanks{$^*$Y. Du and G. Chen are co-corresponding authors of this paper (email: \textit{dy020@ie.cuhk.edu.hk}, \textit{gychen@link.cuhk.edu.hk}).}}
    \vspace{-2.5em}
}

\maketitle

\begin{abstract}
The development of AI-assisted chemical synthesis tools requires comprehensive datasets covering diverse reaction types, yet current high-throughput experimental (HTE) approaches are expensive and limited in scope. Chemical literature represents a vast, underexplored data source containing thousands of reactions published annually. However, extracting reaction information from literature faces significant challenges including varied writing styles, complex coreference relationships, and multimodal information presentation. This paper proposes ChemMiner, a novel end-to-end framework leveraging multiple agents powered by large language models (LLMs) to extract high-fidelity chemical data from literature. ChemMiner incorporates three specialized agents: a text analysis agent for coreference mapping, a multimodal agent for non-textual information extraction, and a synthesis analysis agent for data generation. Furthermore, we developed a comprehensive benchmark with expert-annotated chemical literature to evaluate both extraction efficiency and precision. Experimental results demonstrate reaction identification rates comparable to human chemists while significantly reducing processing time, with high accuracy, recall, and F1 scores. Our open-sourced benchmark facilitates future research in chemical literature data mining.
\end{abstract}


\section{Introduction} \label{sec:intro}

\begin{figure*}[htbp]
\centering
\includegraphics[width=1.0\textwidth]{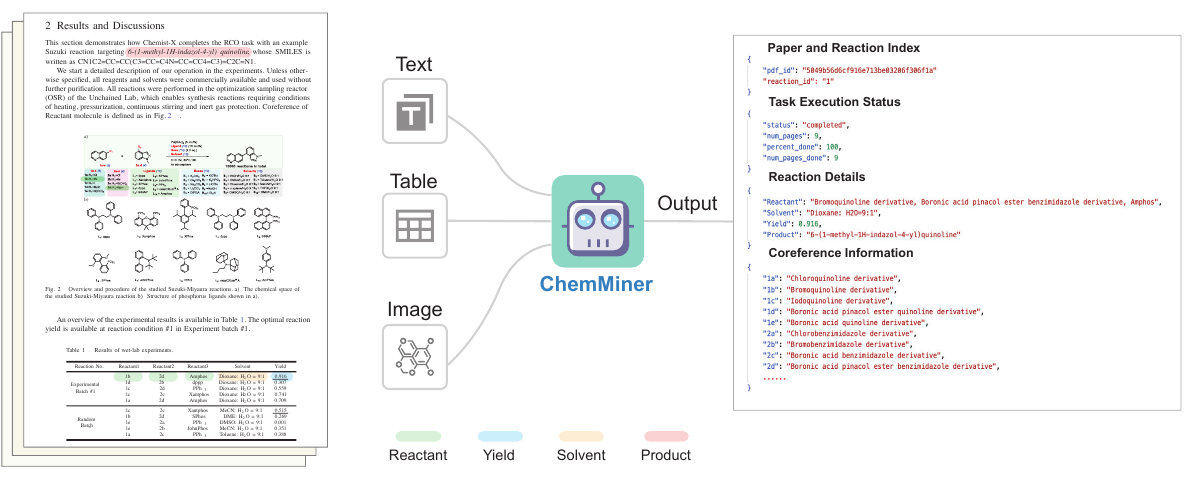}
\caption{An illustration of the reaction extraction task. According to the example literature, three reactants are involved in the synthesis of the desired product (highlighted in red in the given literature). Due to lengthy molecule names, the first two reactants are coreferenced as \textit{1b} and \textit{2d} (information available in the image of the example literature, highlighted in green), while the third reactant is directly named as ``Amphos" (information available in the table, also highlighted in green). Information about the solvent required and the final reaction yield is given in the table, highlighted in yellow and blue, respectively. This example shows the common existence of coreference in chemical literature. Meanwhile, it emphasizes the necessity of multimodal documentary processing ability for the extraction of reaction information - the system must simultaneously analyze the text, image, and table within the given literature to obtain the desired reaction information (i.e., reactants, solvent, product, and yield in this example).}\label{teaser}
\end{figure*}

The discipline of chemistry, characterized by its immense potential and practical utility, is deeply intertwined with the synthesis of materials and the discovery of drugs. In recent studies \cite{a7, a8}, AI-empowered synthesis tools have demonstrated the potential to assist human chemists in precisely identifying reaction patterns and optimizing novel synthetic routes. 

Despite the early-stage successes of AI-assisted tools in chemical synthesis, the continual evaluation of such domain-specific AIs requires a substantial amount of data, which enables the AI to better understand the underlying reaction principles through model training \cite{c1, c2, c3}. As a result, building a comprehensive synthesis dataset that covers a wide variety of chemical reaction types has become a critical challenge within the research community.

A mainstream approach to dataset construction, as represented by \cite{b4, b7, lu2024roboticized}, is to conduct high-throughput experiments (HTEs) for data collection. However, the high expense of HTE has limited its wide application as a data collection method. Moreover, a group of HTEs typically covers only one type of chemical reaction, i.e., relying on automatic chemical platforms to test different reaction conditions and reactant combinations without changing the reaction type. For the training of chemical AI models, however, a diverse training dataset with different reaction types is more preferable.

Chemical literature, on the other hand, remains a largely unexplored data source with rich information diversity. Thousands of chemical paper get published every year, with a massive amount of reaction of different types being conducted to justify the research outcomes. An efficient utilization of these previously published papers can significantly enlarge the reaction dataset with a relatively small cost.

However, extracting reaction information from intricate literature presents several technical challenges. In scientific literature, authors with different technical backgrounds and writing styles may describe their chemical reactions in completely different manners, posing significant challenges for conventional rule-based pattern recognition methods in text analysis. Furthermore, given the lengthy representation of chemical items, molecule coreferences, in which the combination of a simple number and an English letter, such as ``Item 1a”, is used as the abbreviation for a very complex organic molecule name, are frequently used. This form of shorthand, while effective for human readers familiar with the context, presents a unique challenge for machine-reading and understanding. Therefore, precise literature analysis and information extraction of chemical descriptions full of component coreferences remains problematic. Additionally, information in scientific papers is often presented in a multimodal manner for the reader's easier understanding, such as using figures to show reaction conditions and tables to list synthesis yields. This multimodal presentation creates additional challenges for document analysis. Fig. \ref{teaser} gives an example for the reaction information extraction task, in which both the coreferences issue and the multimodal issue are well illustrated. This example also illustrates the expected outcome of reaction extraction task.

Traditional methods for literature information analysis, such as manual extraction, rule-based extraction, and machine learning methods, are insufficient for addressing these challenges \cite{d1, d2, a5}. Manual extraction relies on the intensive labor of chemists, which escalates the cost of the dataset and limits its size. Rule-based methods struggle to adapt to new chemical literature with different writing styles. Machine learning frameworks, such as the one proposed in \cite{a5}, are hampered by the scarcity of annotated reactions, subsequently diminishing the model's performance.

\begin{figure*}[htbp]
\centering
\includegraphics[width=1.0\textwidth]{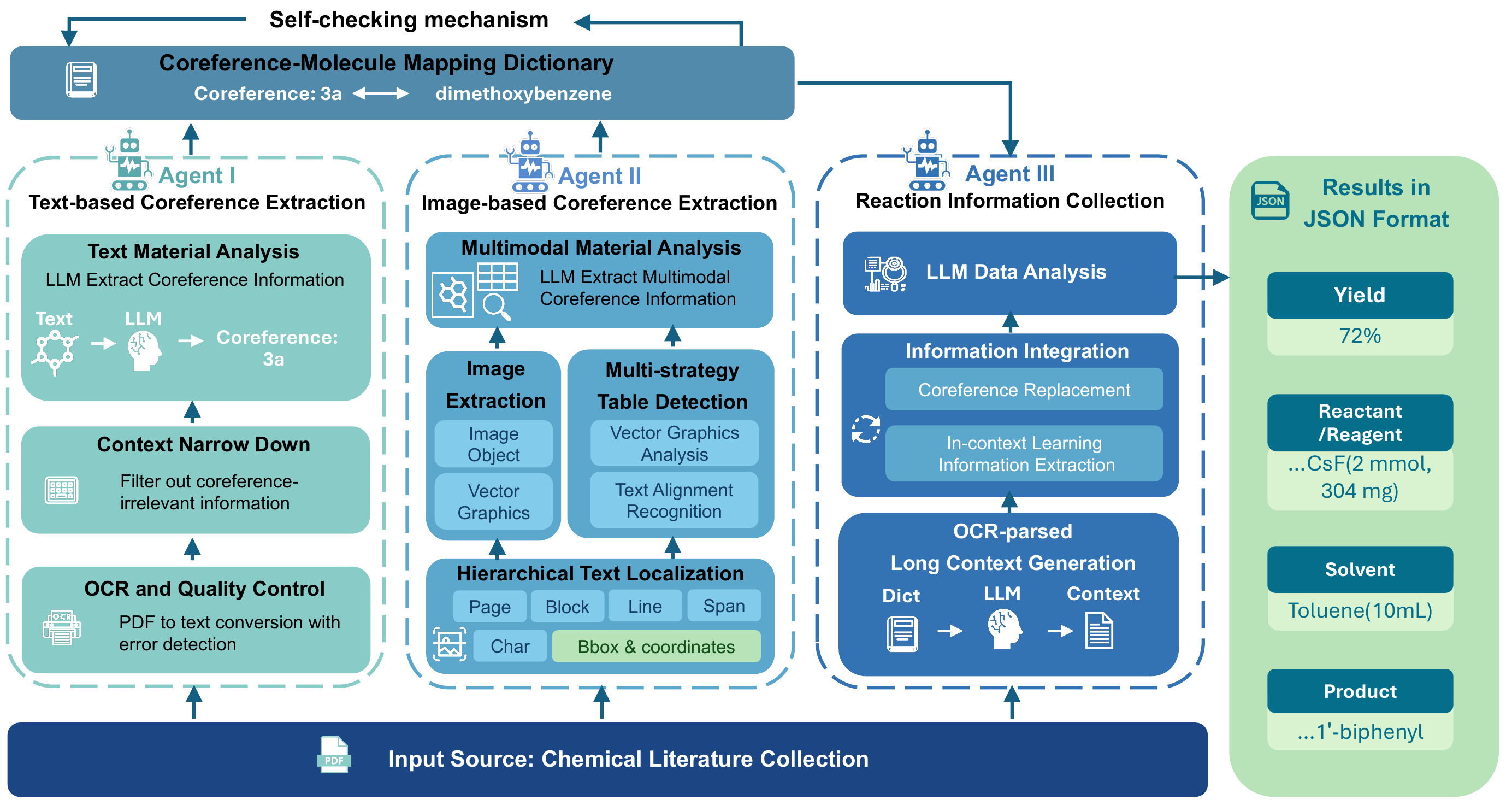}
\caption{The framework of chemical literature analysis and reaction information extraction system based on LLMs.}\label{fig1}
\end{figure*}

The introduction of large language models (LLMs) in data mining research \cite{a9, a10} has provided a new possibility for constructing literature-based chemical reaction datasets. This paper proposes ChemMiner, an end-to-end framework that leverages the cooperation of LLM-empowered AIs to extract high-fidelity chemical data from literature. The framework, as shown in Fig. \ref{fig1}, incorporates three independent agents for the extraction task. To begin with, we deploy a text analysis agent after the textual information of the paper has been obtained with an optical character recognition (OCR) module. The task of this agent is to maintain a dictionary that records the mapping between chemical components and their coreferences appearing in the paper. The second agent addresses the coreference relationships appearing in modalities other than simple text, e.g., tables or figures. This agent is designed to simplify the task of the first agent, and the coreference information it extracts serves as a supplement to the first agent's outcome. With the resulting dictionary maintained by the first two agents, the third agent is responsible for summarizing the literature and generating data instances.

To evaluate the proposed ChemMiner framework, especially its performance for the unique coreference problem in chemical literature, we developed a comprehensive testing benchmark that includes both text-based reaction extraction tasks and multimodal-based ones. Literature tested in this dataset was manually annotated by human chemistry experts to ensure a reliable ground truth. With this testing benchmark, we first evaluate the \textit{efficiency} of ChemMiner's data extraction process. Specifically, we compare the multi-agent system with a human chemist and evaluate the performance from two aspects: 1) the proportion of successfully identified chemical reactions; 2) the time consumption in the reaction extraction process. Meanwhile, we are also interested in the \textit{accuracy} of identified reactions. To this end, we analyzed the accuracy, recall, and F1 score of those successfully extracted reactions.

Our main contributions are summarized as follows:
\begin{itemize}
\item \textbf{Problem Identification}: We identified a critical research problem in chemical literature data mining and highlighted two technical challenges to be addressed in this new task - the coreference issue and multimodal document analysis.
\item \textbf{Framework Design and System Implementation}: We developed ChemMiner, a novel framework that addresses these challenges in chemical literature data mining through the cooperation of multiple LLM agents. We implemented this multi-agent system for comprehensive testing. Experimental results of ChemMiner validated the efficiency and precision of the proposed mechanism. %as well as the effectiveness of each individual element within this framework.
\item \textbf{Dataset and Benchmark}: We collected a chemist-annotated testing benchmark during the evaluation process. This high-quality dataset tests model performance in chemical literature data mining in a multimodal manner and contains valuable information such as a human chemist's time consumption for given literature. We have open-sourced the testing benchmark to facilitate future research in this direction.\footnote{Data and code will be made public upon the acceptance of this paper.}
\end{itemize}

\begin{figure*}[htbp]
\centering
\includegraphics[width=0.71\textwidth]{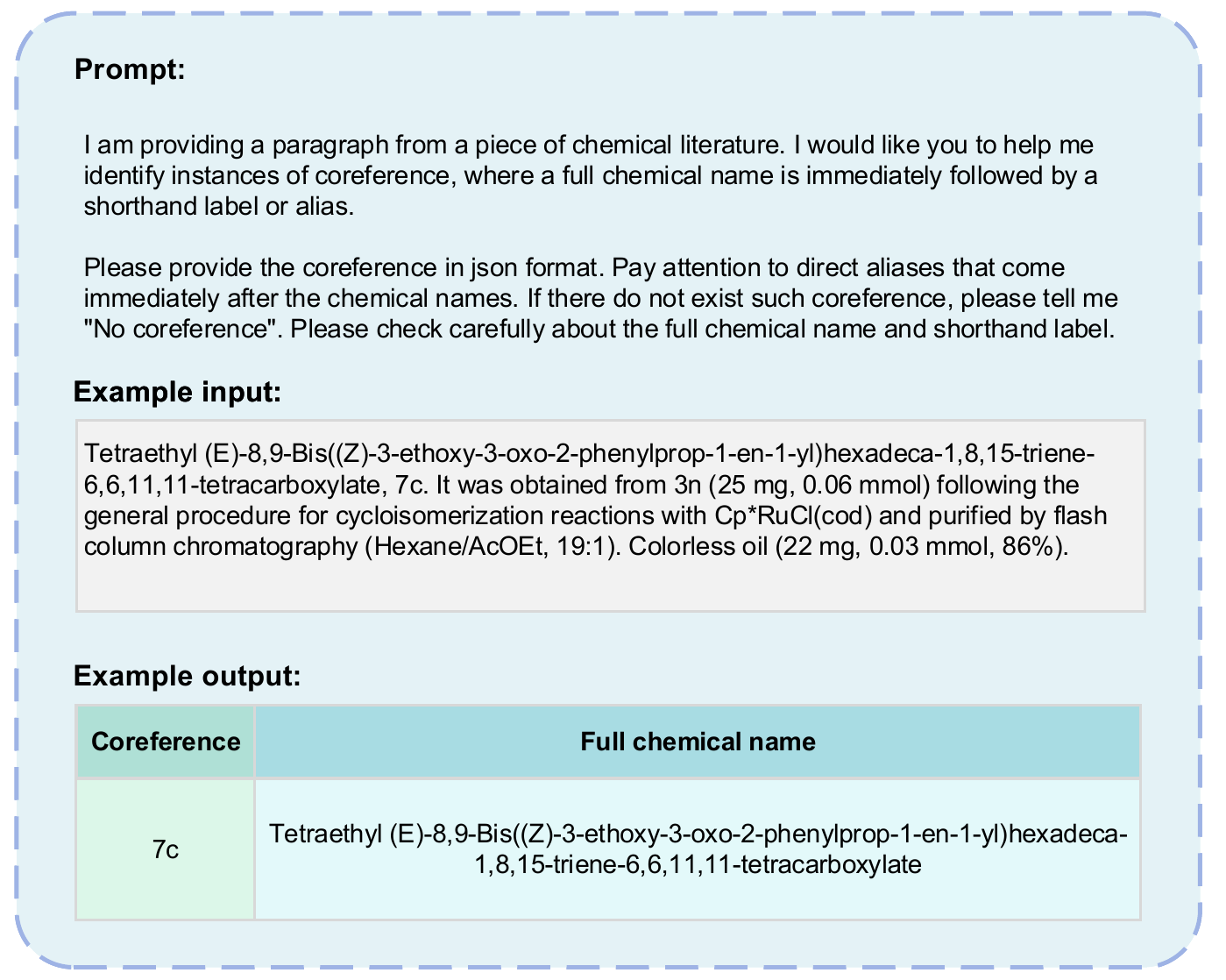}
\caption{The prompt, example input, and example output of the coreference identification task in Agent I.}\label{fig2}
\end{figure*}
\section{Methodology} \label{sec:method}

As illustrated in Fig. \ref{fig1}, ChemMiner has four major components -- three agents of different functionalities and a shared dictionary recording the coreference-molecule mapping relationships. The following of this section gives a detailed introduction to the design and implementation of each individual agent.

\subsection{Agent I: Text-based Coreference Extraction}
The task of Agent I is to extract the coreference information from the given textual material. Since a cross-page coreference is a very rare case, we segment a research paper, typically in the form of a PDF file, into multiple individual pages and ask the agent to process them one by one. 

To convert the content in the given PDF page into a machine-readable format, we conduct OCR as the first step of Agent I's processing. The OCR process, while generally accurate, is not infallible~\cite{gooding2013myth}. It occasionally struggles with complex layouts and low-quality scans, leading to potential errors in the resulting text. To address these possible errors, we also implemented a quality control mechanism along with the OCR module to ensure reliable inputs for the subsequent procedures. The mechanism is, in essence, a sampling-based error detection process. Specifically, we used key phrases frequently appeared in chemical papers, such as ``General Procedure", ``Typical Procedure", or ``General Experiment", as indicators of the OCR quality. Post-OCR materials~\cite{nguyen2021survey} that misspell these common phrases were deemed to be of insufficient quality and were thus marked as mistaken ones. In practical applications, those rare cases can be manually edited to eliminate errors therein.

After OCR, we narrow down the content that awaits analysis to filter out coreference-irrelevant information. Typically, molecule coreferences appear in technical sections such as the methodology or the experimental sections. Therefore, once the agent identifies that the current content does not belong to a technical section, it bypasses the subsequent analysis of the current content and moves on to the next page to enhance the efficiency of coreference extractions.

\begin{figure*}[htbp]
\centering
\includegraphics[width=0.76\textwidth]{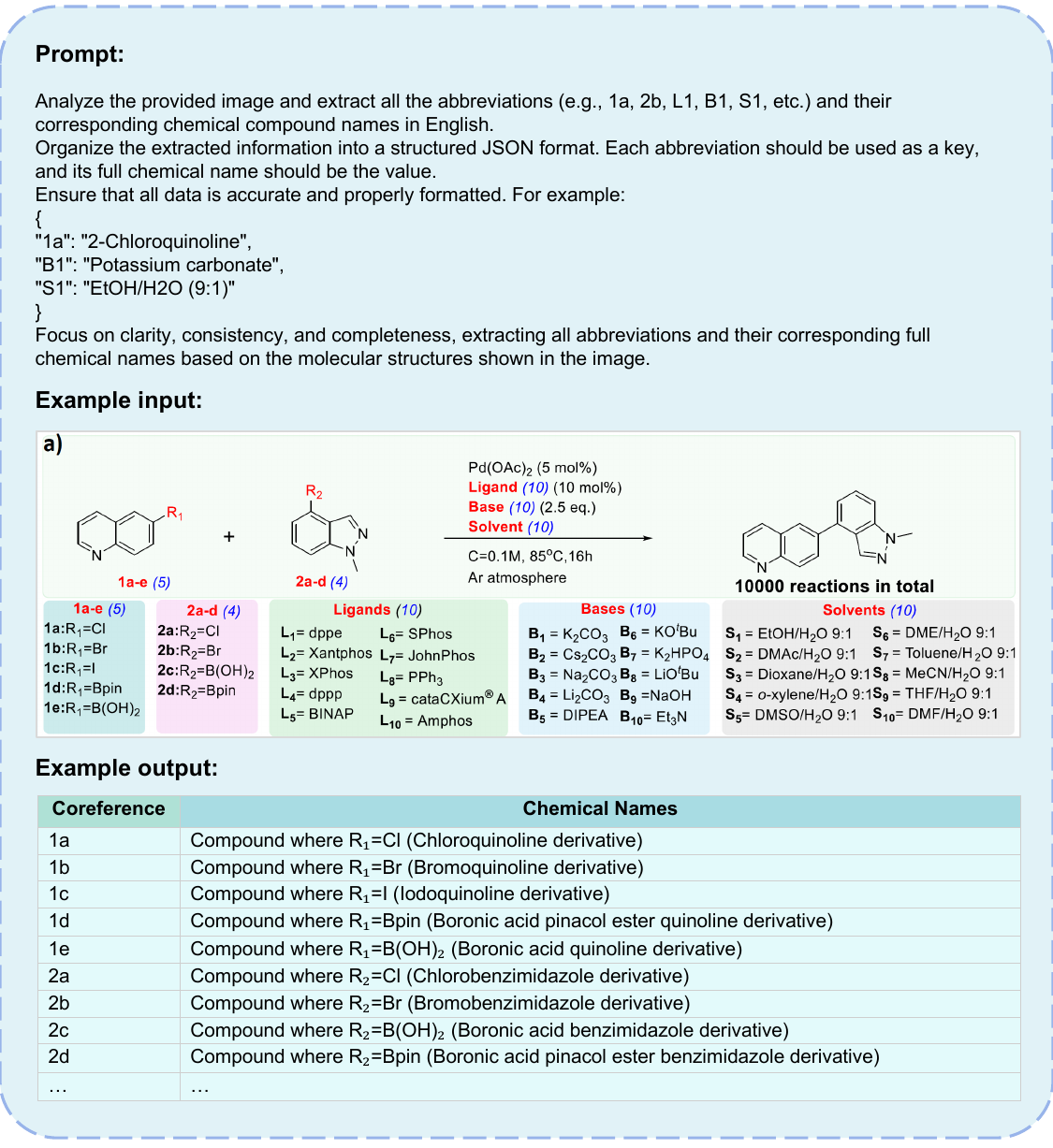}
\caption{The prompt, example input, and example output of the multimodal coreference identification task in Agent II.}\label{fig3}
\end{figure*}

\begin{figure*}[htbp]
\centering
\includegraphics[width=0.73\textwidth]{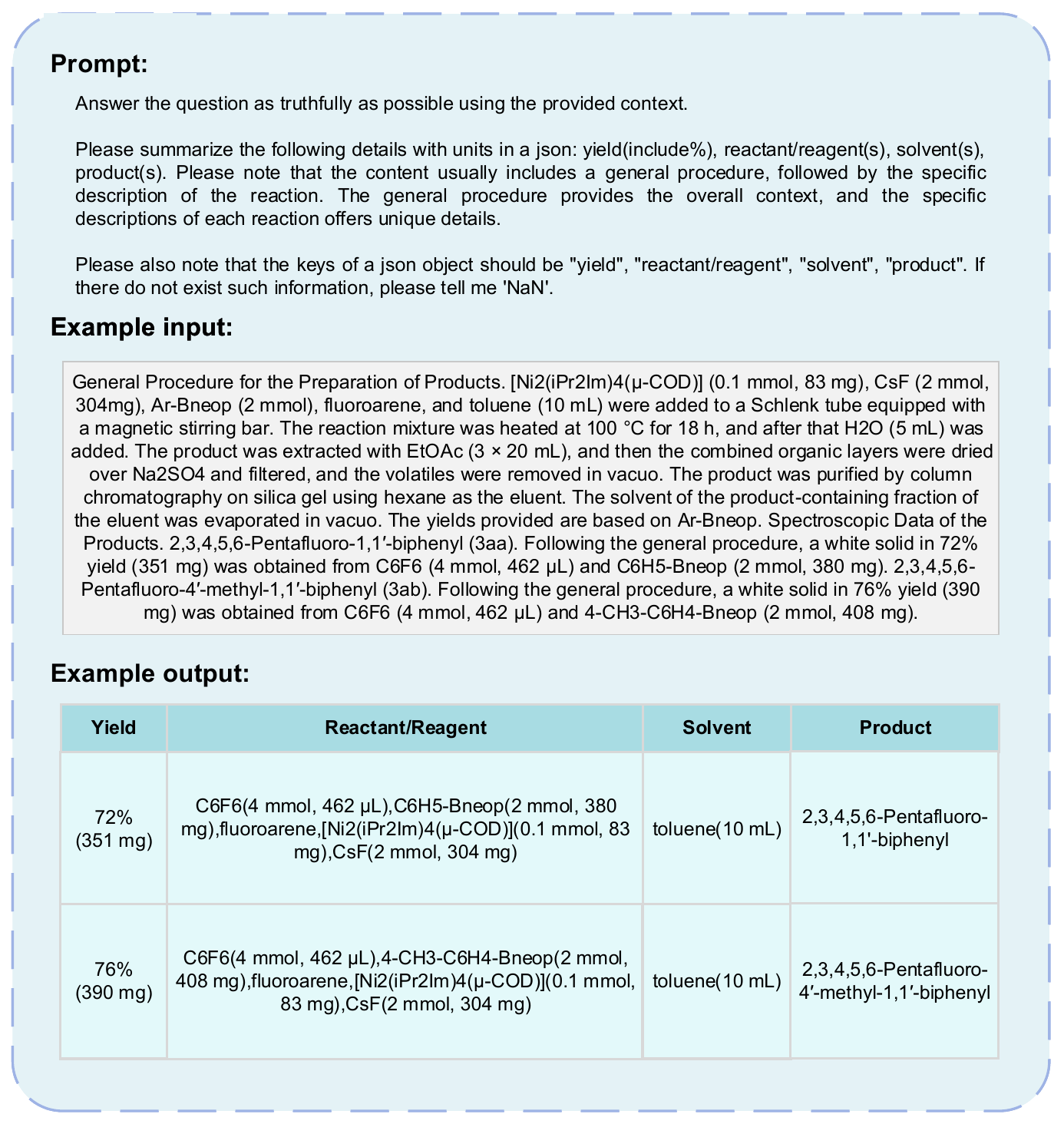}
\caption{The prompt, example input, and example output of chemical reaction information extraction task in Agent III.}\label{fig4}
\end{figure*}

Finally, we use the textual analysis ability of LLMs to extract the coreference information within the given text. Fig. \ref{fig2} presents the prompt, example input, and example output of the in-context learning \cite{wang2025cellular} process in coreference extraction. Upon encountering a potential coreference, the AI Agent validates it against the patterns typically used for coreferences in the chemical literature. This step ensures that the identified coreferences are not false positives, such as a number and a letter appearing together by coincidence in the text. Once a coreference is validated, the AI Agent records it for subsequent steps.

\subsection{Agent II: Image-based Coreference Extraction}

The task of Agent II is to extract the molecule coreference from given PDF document pages, with a focus on the information presented in the form of images and tables. 

To obtain coreference information from images and tables within a PDF file, we need to extract the associated images and tables first. To this end, we decode and process the PDF file, converting it into a structured hierarchical format with five levels: pages, blocks, lines, spans, and characters. Each character retains precise coordinates and bounding box information for accurate positioning. To assist the extraction of images and tables, text content is sorted by vertical and then horizontal coordinates to generate natural reading sequences.

With the textual information obtained, the agent performs table detection using a multi-strategy fusion approach. The system combines vector graphics analysis for tables with explicit borders and text alignment pattern recognition for implicit table structures. The text alignment detection simultaneously extracts the left and right boundary coordinates, as well as the center point coordinates of words, and uses a predefined clustering tolerance to group spatially proximate words into alignment groups, similar to prior established works~\cite{reddy2016text}. This generates virtual boundary lines to reconstruct table structures from spatial distribution patterns. The detected boundaries are then optimized through line merging and intersection calculations to construct table grids. Table segments from adjacent pages are merged into complete tables.

Image extraction adopts different ways to handle different image types in PDFs. For embedded XObject image resources in PDFs, the system directly extracts original data while preserving native format and resolution. For vector graphic content, the system performs rasterized rendering and clipping to convert it into bitmap format. This approach maximizes quality preservation while optimizing processing performance.

After the extraction of images and tables, the agent uses a multimodal LLM for material analysis. Fig. \ref{fig3} gives an example for the process, with detailed prompting information, example input, and example outputs presented for the reader's easier understanding.

\subsection{Coreference-Molecule Mapping Dictionary}
The aim of Agents I and II is to establish the precise mapping between a coreference appearing in the given literature and the associated molecule. For that purpose, we implemented a dictionary to record the coreference-molecule relationship found by Agent I and Agent II. Furthermore, the dictionary also has a self-checking mechanism - when a new coreference-molecule pair is added, the dictionary will check if this new pair is duplicated or contradicts existing mapping relationships stored. If any of these problems happen, the agent system calls for a revisit of the coreference-molecule pair involved to ensure the reliability of the information collected.

\subsection{Agent III: Reaction Information Collection}
Building upon the coreference-molecule mapping relationship established, Agent III aims to summarize the chemical information within the given article and store them in a JSON file for later examinations. In the information extraction process, we are interested in the yield, reactant, catalyst, solvent, and products of each chemical reaction identified. 

Different from the setup in Agent I, in which the LLM is required to find out the mapping between a molecule's coreference name and full name, the information extraction in Agent III needs to be conducted in a more comprehensive way given that the wanted information of a reaction might be given in different parts of a paper. For example, a paper may introduce the yield of investigated reaction in abstract or introduction of the paper to attract readers' interest, while detailed reaction conditions, such as catalyst and solvent used, are given in the experimental sections later. That calls for better long-content processing ability for the implementation of Agent III. In this paper, we realized Agent III with GPT4-Turbo, which has the ability to process 128k tokens in one prompt, while the implementation of Agent I was based on GPT4, which accepts 32k tokens only.

With the long content obtained with the whole PDF file, an comprehensive integration is conducted within Agent III. The agent uses the in-context learning framework described in Fig. \ref{fig4} to extract chemical reaction information from the post-OCR textual material. And then, based on the coreference-molecule dictionary constructed by the first two agents, Agent III replaces all coreferences in the identified reaction. This is followed by a systematic summarization and structuralization of the obtained information. The resulting data, such as the one shown in Fig. \ref{fig4}, are stored in a pre-defined JSON format (see Fig. \ref{teaser} for an example of the JSON format).
\section{Experimental Results} \label{sec:experiments}

Acting as an efficient helper for chemists in the data mining task, ChemMiner is expected to identify reaction information from literature in an efficient and error-free manner. This section quantitatively measures ChemMiner's ability in terms of efficiency and precision.

\begin{figure}[htbp]
\centering
\includegraphics[width=0.475\textwidth]{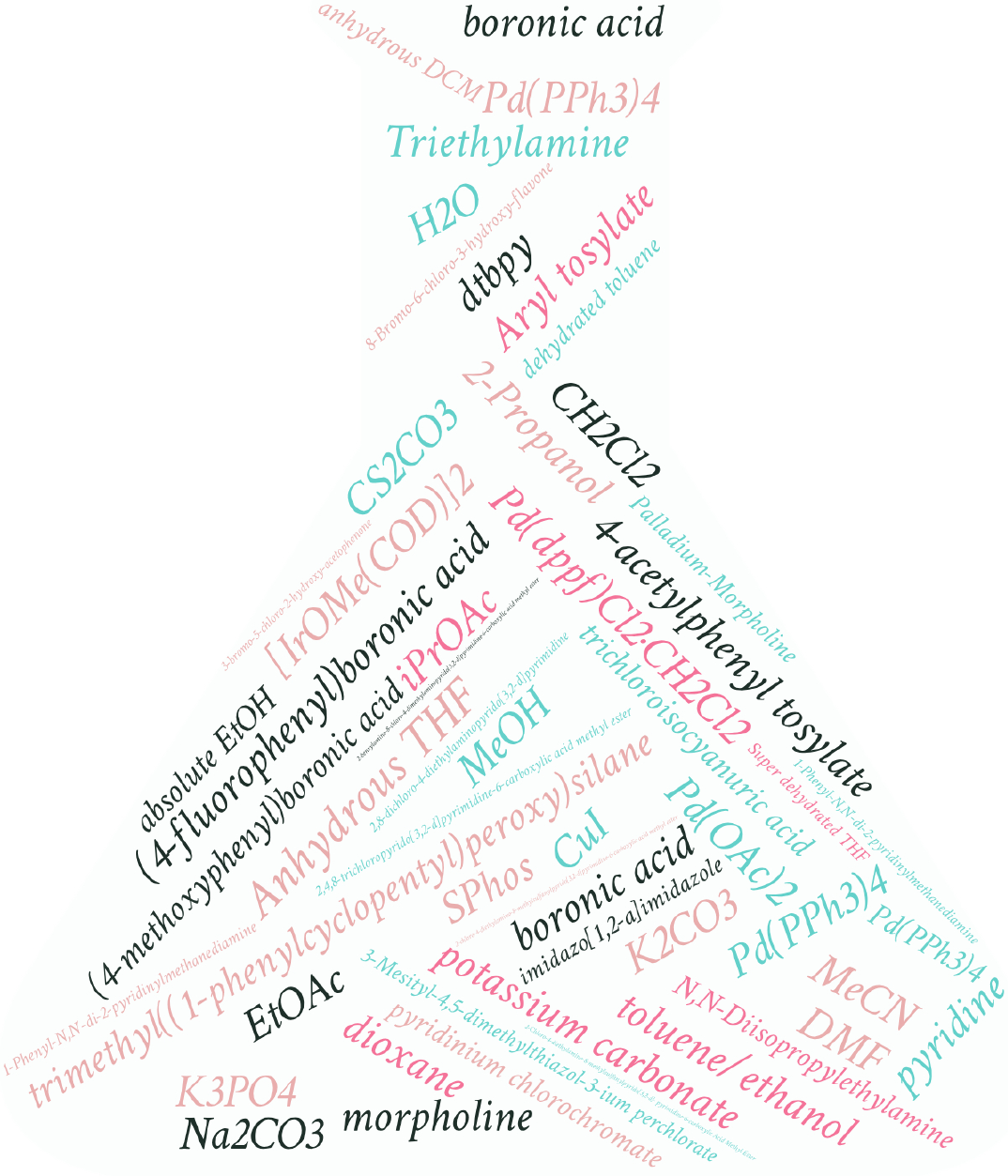}
\caption{A visualization of the dataset distribution, presented in the shape of a chemical flask to reflect the chemistry theme.}\label{fig6}
\end{figure}

We first introduce the testing dataset we have established. For the building of this benchmark, we collected 1940 chemical research papers from open-source literature platforms. We plot a word cloud in Fig. \ref{fig6} to visualize the data distribution within this literature dataset.

Since the baseline aims to provide an equal evaluation for systems developed to address this task, we pre-processed the collected PDF research papers with OCR and the image/table identification and extraction pipeline (see the illustration in Fig. \ref{fig1} and the discussions in Section \ref{sec:method} for details). This is to make sure the evaluation can better focus on the core documentary analysis part, i.e., eliminating the potential performance differences brought by different implementations of these peripheral preprocessing steps. The pre-processed data is available in the project's github repository.

We also performed manual labeling for the literature data. In total 17 randomly selected papers have been proofread and annotated by human chemists -- all reaction-related information (e.g., yield, reactant, catalyst, solvent, and products of a reaction) are manually collected. Note that we did not label all chemical literature in the dataset due to limited time and manpower in the chemical field. With the further investigation of our newly proposed research problem, we will continue the labeling as continual contributions to the community. The following evaluation is based on the 326 reaction information obtained from the textual description and image/table description in those selected papers.

\begin{table}[!ht]
    \caption{The number of correct/extracted/total reactions}
    \centering
    \begin{tabular}{lccc}
    \toprule
        ~ & Correct data & Extracted data & Total data \\ 
    \midrule
        Yield & 236 & 256 & 326 \\ 
        Reactant & 203 & 228 & 300 \\ 
        Solvent & 227 & 247 & 326 \\ 
        Product & 223 & 255 & 326 \\ 
    \bottomrule
    \end{tabular} \label{Table1}
\end{table}

We start from the evaluation of data mining efficiency. With the focus on the yield, reactant, solvent, and products of each reaction, Table \ref{Table1} presents the following figures: 1) the total number of reactions identified from the given literature as ground truth\footnote{A common phenomenon in chemical literature is that some papers may not give the full list of required reactants for classic reactions familiar to its readers \cite{e1, e2, e3}. While chemical experts know the missing information about the reactants, the reactant information should not be treated as ground truth for the evaluation of a data mining process. Therefore, in Table \ref{Table1}, the ground truth number of available reactants is smaller than 326.}; 2) the number of reactions that have been successfully extracted by ChemMiner; and 3) the amount of reaction information that has been correctly extracted (i.e., we compare the yield, reactant, solvent, and products of an extracted reaction with the ground truth).

From Table \ref{Table1}, we can calculate the ratio of correctly extracted reaction information, which represents the efficiency of the agent system. Specifically, this efficiency value is 72.39\%, 67.67\%, 69.63\%, and 68.40\% for yield, reactant, solvent, and product, respectively.

\begin{table}[!ht]
    \caption{Precision, recall, and F1 score of the data mining results}
    \centering
    \begin{tabular}{lccc}
    \toprule
        ~ & Precision & Recall & F1-score \\ 
    \midrule
        Yield & 92.19\% & 78.53\% & 84.81\% \\ 
        Reactant & 89.04\% & 76.00\% & 82.00\% \\ 
        Solvent & 91.90\% & 75.77\% & 83.06\% \\ 
        Product & 87.45\% & 78.22\% & 82.58\% \\ 
    \bottomrule
    \end{tabular} \label{Table2}
\end{table}

\begin{table*}[!ht]
    \caption{The precision, average cost, and average speed of manual data collection and our agent.}
    \centering
    \begin{tabular}{lccc}
    \toprule
        ~ & Precision & Time Consumption per Reaction (second) & Cost per Reaction (USD) \\ 
    \midrule
        Manual Data Collection & 90.05\% & 288 & 1.41 \\ 
        AI Agent & 87.45\%  & 0.43 &  0.0025 \\ 
    \bottomrule
    \end{tabular} \label{Table3}
\end{table*}

We then look at the accuracy of the reaction extraction process. In Table \ref{Table2}, we consider the information extraction result as a binary classification problem (i.e., correctly and incorrectly identified) and calculate the precision, recall, and F1 score for the four types of information of interest to us - reactant, solvent, yield, and the resulting products. On average, the precision, recall, and F1-score of the information identification reach 90.15\%, 77.13\%, and 83.11\%, respectively.

Finally, in terms of system benchmarking, there are no other open-source tools tailored to the extraction of reaction data from chemical journals according to our best knowledge, given that this is a new task never investigated in previous work. Nevertheless, we can validate the performance of the agent through comparison with the manual data collection process of human experts. 

To this end, we gave those selected papers to six junior level postgraduate students majoring in chemistry and asked them to do the reaction data extraction task manually. Note that dictionary and Internet are allowed in the experiment to simulate a real-world working scenario, but AI tools are forbidden. After they have submitted their answers, we compared their results with the ground truth annotated by a research scientist in chemistry to find the precision of each answer. During the test, we also recorded the time consumption of each student, and calculated the payment for each student according to their working hours and the standard hourly payment for postgraduate-level student helpers.\footnote{To avoid a student's subjective bias on the time consumption, we did not tell them that they would be paid according to their working hours until the end of experiments.}

Table \ref{Table3} benchmarks ChemMiner with human students from three aspects: the average precision of reaction identification, 2) the average time required for the identification of reaction, and 3) the average cost for identifying one reaction. As we can see from the table, our agent system achieves a comparable performance as an average human student does, while it has significantly lower consumption and financial cost.
\section{Conclusion and Future Work} \label{sec:conclusion}
This paper presented ChemMiner, a novel multi-agent framework leveraging LLMs for automated extraction of chemical reaction information from scientific literature. The framework addresses two critical challenges in chemical literature data mining: the complex coreference problem where simple abbreviations represent complex molecular structures, and the multimodal nature of information presentation in scientific documents.

ChemMiner demonstrates significant promise in automating the traditionally labor-intensive process of chemical data extraction. The experimental evaluation on a testing benchmark of chemical research papers reveals encouraging results, with extraction efficiency rates ranging from 67.67\% to 72.39\% across different types of chemical information. The framework achieves strong performance with an average precision of 90.15\%, recall of 77.13\%, and F1-score of 83.11\%. Notably, the system achieves comparable accuracy to human chemistry students while requiring significantly less time and financial cost.

The framework's ability to handle the unique challenges of chemical literature -- particularly the pervasive use of coreferences and multimodal information presentation -- represents a significant advancement in automated chemical data mining. This work establishes a foundation for building comprehensive reaction datasets from the vast repository of published chemical literature, potentially accelerating AI-assisted chemical synthesis research.

Future work can focus on several key directions. Expanding the evaluation dataset with more comprehensive manual annotations would strengthen validation across diverse chemical domains. Enhanced error detection and correction mechanisms could further improve extraction accuracy through cross-validation between agents or uncertainty quantification. The framework could be extended to handle additional reaction parameters such as temperature, pressure, and reaction time. Integration of domain-specific chemical knowledge bases could enhance understanding and resolve ambiguities. Advanced machine learning techniques such as active learning could be incorporated to continuously improve extraction performance with minimal human supervision. Additionally, developing real-time processing capabilities would allow integration with ongoing research workflows, enabling immediate data extraction as new literature becomes available. Cross-language support could expand the framework's applicability to global chemical literature, while specialized handling of patent documents could unlock valuable proprietary reaction data. Finally, developing standardized evaluation metrics and benchmarks for chemical literature data mining would benefit the broader research community.

\bibliographystyle{IEEEtran}
\bibliography{achemso-demo}

\begin{thebibliography}{10}
\providecommand{\url}[1]{#1}
\csname url@samestyle\endcsname
\providecommand{\newblock}{\relax}
\providecommand{\bibinfo}[2]{#2}
\providecommand{\BIBentrySTDinterwordspacing}{\spaceskip=0pt\relax}
\providecommand{\BIBentryALTinterwordstretchfactor}{4}
\providecommand{\BIBentryALTinterwordspacing}{\spaceskip=\fontdimen2\font plus
\BIBentryALTinterwordstretchfactor\fontdimen3\font minus \fontdimen4\font\relax}
\providecommand{\BIBforeignlanguage}[2]{{%
\expandafter\ifx\csname l@#1\endcsname\relax
\typeout{** WARNING: IEEEtran.bst: No hyphenation pattern has been}%
\typeout{** loaded for the language `#1'. Using the pattern for}%
\typeout{** the default language instead.}%
\else
\language=\csname l@#1\endcsname
\fi
#2}}
\providecommand{\BIBdecl}{\relax}
\BIBdecl

\bibitem{a7}
K.~Chen, G.~Chen, J.~Li, Y.~Huang, E.~Wang, T.~Hou, and P.-A. Heng, ``{MetaRF}: attention-based random forest for reaction yield prediction with a few trails,'' \emph{Journal of Cheminformatics}, vol.~15, no.~1, pp. 1--12, 2023.

\bibitem{a8}
K.~Chen, J.~Li, K.~Wang, Y.~Du, J.~Yu, J.~Lu, G.~Chen, L.~Li, J.~Qiu, Q.~Fang \emph{et~al.}, ``Towards an automatic ai agent for reaction condition recommendation in chemical synthesis,'' \emph{arXiv preprint arXiv:2311.10776}, 2023.

\bibitem{c1}
Z.~Zheng, N.~Rampal, T.~J. Inizan, C.~Borgs, J.~T. Chayes, and O.~M. Yaghi, ``Large language models for reticular chemistry,'' \emph{Nature Reviews Materials}, pp. 1--13, 2025.

\bibitem{c2}
Y.~Zimmermann, A.~Bazgir, A.~Al-Feghali, M.~Ansari, J.~Bocarsly, L.~C. Brinson, Y.~Chiang, D.~Circi, M.-H. Chiu, N.~Daelman \emph{et~al.}, ``34 examples of llm applications in materials science and chemistry: Towards automation, assistants, agents, and accelerated scientific discovery,'' \emph{arXiv preprint arXiv:2505.03049}, 2025.

\bibitem{c3}
Z.~Zhao, D.~Ma, L.~Chen, L.~Sun, Z.~Li, Y.~Xia, B.~Chen, H.~Xu, Z.~Zhu, S.~Zhu \emph{et~al.}, ``Developing chemdfm as a large language foundation model for chemistry,'' \emph{Cell Reports Physical Science}, vol.~6, no.~4, 2025.

\bibitem{b4}
D.~T. Ahneman, J.~G. Estrada, S.~Lin, S.~D. Dreher, and A.~G. Doyle, ``Predicting reaction performance in c--n cross-coupling using machine learning,'' \emph{Science}, vol. 360, no. 6385, pp. 186--190, 2018.

\bibitem{b7}
D.~Perera, J.~W. Tucker, S.~Brahmbhatt, C.~J. Helal, A.~Chong, W.~Farrell, P.~Richardson, and N.~W. Sach, ``A platform for automated nanomole-scale reaction screening and micromole-scale synthesis in flow,'' \emph{Science}, vol. 359, no. 6374, pp. 429--434, 2018.

\bibitem{lu2024roboticized}
J.-M. Lu, H.-F. Wang, Q.-H. Guo, J.-W. Wang, T.-T. Li, K.-X. Chen, M.-T. Zhang, J.-B. Chen, Q.-N. Shi, Y.~Huang \emph{et~al.}, ``Roboticized ai-assisted microfluidic photocatalytic synthesis and screening up to 10,000 reactions per day,'' \emph{Nature Communications}, vol.~15, no.~1, pp. 1--13, 2024.

\bibitem{d1}
J.~Zhang, K.~C. Tsui, H.~Y. Lee, L.~Aquili, K.~H. Wong, E.~Kocabicak, Y.~Temel, Z.~Lu, M.-L. Fung, A.~Kalueff \emph{et~al.}, ``Data mining approach to melatonin treatment in alzheimer’s disease: New gene targets mmp2 and nr3c1,'' \emph{International Journal of Molecular Sciences}, vol.~26, no.~1, p. 338, 2025.

\bibitem{d2}
M.~Schilling-Wilhelmi, M.~R{\'\i}os-Garc{\'\i}a, S.~Shabih, M.~V. Gil, S.~Miret, C.~T. Koch, J.~A. M{\'a}rquez, and K.~M. Jablonka, ``From text to insight: large language models for chemical data extraction,'' \emph{Chemical Society Reviews}, 2025.

\bibitem{a5}
J.~Guo, A.~S. Ibanez-Lopez, H.~Gao, V.~Quach, C.~W. Coley, K.~F. Jensen, and R.~Barzilay, ``Automated chemical reaction extraction from scientific literature,'' \emph{Journal of chemical information and modeling}, vol.~62, no.~9, pp. 2035--2045, 2021.

\bibitem{a9}
H.~Cui, Y.~Du, Q.~Yang, Y.~Shao, and S.~C. Liew, ``{LLMind}: Orchestrating {AI} and {IoT} with {LLMs} for complex task execution,'' \emph{arXiv preprint arXiv:2312.09007}, 2023.

\bibitem{a10}
Y.~Du, S.~C. Liew, K.~Chen, and Y.~Shao, ``The power of large language models for wireless communication system development: A case study on fpga platforms,'' \emph{arXiv preprint arXiv:2307.07319}, 2023.

\bibitem{gooding2013myth}
P.~Gooding, M.~Terras, and C.~Warwick, ``The myth of the new: mass digitization, distant reading, and the future of the book,'' \emph{Literary and Linguistic Computing}, vol.~28, no.~4, pp. 629--639, 2013.

\bibitem{nguyen2021survey}
T.~T.~H. Nguyen, A.~Jatowt, M.~Coustaty, and A.~Doucet, ``Survey of post-ocr processing approaches,'' \emph{ACM Computing Surveys (CSUR)}, vol.~54, no.~6, pp. 1--37, 2021.

\bibitem{wang2025cellular}
L.~Wang, X.~Long, Y.~Du, X.~Liu, K.~Chen, and S.~C. Liew, ``Cellular-x: An llm-empowered cellular agent for efficient base station operations,'' \emph{arXiv preprint arXiv:2504.13190}, 2025.

\bibitem{reddy2016text}
V.~S. Reddy, P.~Kinnicutt, and R.~Lee, ``Text document clustering: the application of cluster analysis to textual document,'' in \emph{2016 International Conference on Computational Science and Computational Intelligence (CSCI)}.\hskip 1em plus 0.5em minus 0.4em\relax IEEE, 2016, pp. 1174--1179.

\bibitem{e1}
V.~Fan, Y.~Qian, A.~Wang, A.~Wang, C.~W. Coley, and R.~Barzilay, ``Openchemie: An information extraction toolkit for chemistry literature,'' \emph{Journal of Chemical Information and Modeling}, vol.~64, no.~14, pp. 5521--5534, 2024.

\bibitem{e2}
A.~Ehnbom and J.~A. Gladysz, ``Gyroscopes and the chemical literature, 2002--2020: approaches to a nascent family of molecular devices,'' \emph{Chemical Reviews}, vol. 121, no.~7, pp. 3701--3750, 2021.

\bibitem{e3}
P.~Chan, T.~Van~Gerven, J.-L. Dubois, and K.~Bernaerts, ``Virtual chemical laboratories: A systematic literature review of research, technologies and instructional design,'' \emph{Computers and Education Open}, vol.~2, p. 100053, 2021.

\end{thebibliography}
\end{document}